\documentclass[letterpaper,twocolumn,10pt]{article}
\usepackage{usenix2022_SOUPS}

\usepackage{tikz}
\usepackage{amsmath}
\usepackage{verbatim}
\usepackage{graphicx}
\usepackage{caption}[singlelinecheck=false,skip=20pt]

\usepackage{subcaption}
\usepackage{multirow}
\usepackage{xurl}
\usepackage{float}

\begin{document}
%-------------------------------------------------------------------------------

%don't want date printed
\date{}

\title{\Large \bf How Well Do My Results Generalize Now? The External Validity of \mbox{Online Privacy and Security Surveys}} 

% if you leave this blank it will default to a possibly ugly attempt 
% to make the contents of the \author command below into a string
\def\plainauthor{Jenny Tang, Eleanor Birrell, and Ada Lerner}
%\def\plainauthor{Author(s) anonymized for review.}

%for single author (just remove % characters)
%\author{Author(s) anonymized for review}
%\begin{comment}
\author{
{\rm Jenny Tang}\\
Wellesley College
\and
{\rm Eleanor Birrell}\\
Pomona College
\and
{\rm Ada Lerner\thanks{Work was done while Lerner was at Wellesley College.}}\\
Northeastern University
} % end author
%\end{comment}

%\pagenumbering{gobble} % suppress page numbers
\maketitle
%\thecopyright % commented out for arxiv

%-------------------------------------------------------------------------------
\begin{abstract} % should be under 200 words
%-------------------------------------------------------------------------------
Privacy and security researchers often rely on data collected through online crowdsourcing platforms such as Amazon Mechanical Turk (MTurk) and Prolific. Prior work---which used data collected in the United States between 2013 and 2017---found that MTurk responses regarding security and privacy were generally  representative for people under 50 or with some college education. However, the landscape of online crowdsourcing has changed significantly over the last five years, with the rise of Prolific as a major platform and the increasing presence of bots. This work attempts to replicate the prior results about the external validity of online privacy and security surveys. We conduct an online survey on MTurk ($n=800$), a gender-balanced survey on Prolific ($n=800$), and a representative survey on Prolific ($n=800$) and compare the responses to a probabilistic survey conducted by the Pew Research Center ($n=4272$). We find that MTurk response quality has degraded over the last five years, and our results do not replicate the earlier finding about the generalizability of MTurk responses. By contrast, we find that data collected through Prolific is generally representative for questions about user perceptions and experiences, but not for questions about security and privacy knowledge. We also evaluate the impact of Prolific settings, attention check questions, and statistical methods on the external validity of online surveys, and we develop recommendations about best practices for conducting online privacy and security surveys. 
\end{abstract}

%-------------------------------------------------------------------------------
\section{Introduction}
%-------------------------------------------------------------------------------

Over the last fifteen years, online surveys conducted through crowdsourcing platforms such as Amazon Mechanical Turk (MTurk)~\cite{mturk} and Prolific~\cite{prolific} have become increasingly critical tools for conducting quantitative usable privacy and security research. Researchers often use these platforms to recruit participants in user studies that survey people about topics in privacy and security or evaluate users' interactions with privacy and security tools. However, the external validity of these user studies depends on the extent to which the results of these online studies generalize to the overall population. 

% prior results about quality of mturk/prolific
Prior work has investigated the validity of online surveys in various domains---such as social sciences~\cite{berinsky2014separating, Buhrmester2018AnEO, thompson2020relational}, health behavior~\cite{simons2012common}, and privacy~\cite{kang2014privacy, schnorf2014comparison}---with somewhat mixed results. However, the paper that has arguably been the most influential among the usable privacy and security research community---work by Redmiles et al. based on surveys conducted between 2013 and 2017~\cite{redmiles2019well}---made strong, positive claims about the external validity of privacy and security surveys conducted on MTurk. It found that (1) MTurk responses regarding privacy and security experiences, advice sources, and knowledge were more representative of the U.S. population compared to responses from a census-representative web panel and (2) MTurk responses regarding privacy and security experiences, advice sources, and knowledge were generally representative of the U.S. population for respondents who are younger than 50 or who have some college education. 

% changes to the landscape
However, the landscape of crowdsourcing platforms has changed significantly in the last five years. One key change is the rise of Prolific as a major crowdsourcing platform. Founded in 2014 specifically as a platform for conducting online user studies, Prolific was only rarely used to recruit participants in 2017. By contrast, we find that by 2021, Prolific was approximately twice as common as MTurk as a recruitment platform for usable privacy and security studies. A second key change is the increasing presence of sophisticated bots on MTurk, which can degrade data quality. While bots do not appear to have been a significant problem on MTurk in 2017, more recent work has estimated that 20-50\% of MTurk accounts are actually bots, with significant bot levels dating back to approximately March 2018~\cite{bai2018bots,moss2018after}.

% goal of this work: replication
In light of those changes, this work attempts to replicate the key findings of Redmiles et al.~\cite{redmiles2019well}. We ask:
\begin{itemize}
    \item[(1)] Are MTurk responses to privacy and security survey questions still representative of the U.S. population %for respondents who are younger than 50 or have some college education?
    for respondents under 50 or with some college education?
    \item[(2)] To what extent do various classes of attention check question---reading-based attention checks, open text-response questions, and CAPTCHAs---and/or raking (i.e. demographic weighting) improve the generalizability of MTurk responses?
    \item[(3)] How well do Prolific responses to privacy and security questions generalize to the general U.S. population?
    \item[(4)] What are the current best practices for conducting and analyzing online user surveys in the domain of privacy and security?
\end{itemize}
Additionally, we investigate the limitations of online survey methods for surveying underrepresented demographic groups, reporting on ways that specific groups differ from the general population and how specific populations might be misrepresented by a focus on a representative sample.

% methodology and key results
To answer these research questions, we conduct an online survey on MTurk ($n=800$), a gender-balanced survey on Prolific ($n=800$), and a representative survey on Prolific ($n=800$) and compare the responses to a probabilistic survey conducted through the Pew Research Center ($n=4,272$). We find that MTurk response quality has degraded over the last five years, and our results do not replicate the finding that  MTurk responses are representative of certain subsets of the U.S. population, even when we exclude the 39\% of MTurk responses that fail attention checks and apply raking. We find that data collected through both representative and gender-balanced Prolific samples is generally representative for questions about user experiences, perceptions, and beliefs; however, responses to questions about knowledge of privacy and security concepts and about social media use differ heavily from the overall U.S. population.  We also find that racial, age, and education subgroups from our Prolific representative sample are generally moderately representative of their respective subgroups in the American population. 

% recommendations
Based on our results, we recommend that privacy and security researchers prefer Prolific to MTurk when recruiting participants for online user studies. Our results show that Prolific provides good quality, generalizable data for certain types of user studies about privacy and security (those that focus on experiences, perceptions, and beliefs), but that Prolific users are generally more technical than the overall population, resulting in different responses about knowledge and behavior. We do not recommend using attention check questions or CAPTCHAs on Prolific, as they lengthen surveys unnecessarily without improving external validity.
\section{Related Works}

Given the widespread use of crowdsourcing platforms as recruiting tools for user studies, the question of the data quality and  external validity of online survey data has been extensively studied from a variety of different angles. 

\subsection{Generalizability of Online Platforms}

%Of the user studies published in SOUPS and PETS in the past year, we found that the majority of user studies that involved survey components used Prolific or MTurk. 
Prior work has investigated the generalizability of online user studies conducted through MTurk and Prolific in a variety of different domains.

\paragraph{Amazon Mechanical Turk}
Amazon Mechanical Turk has long been a platform favored by researchers across disciplines such as computer science and the social sciences to conduct user studies~\cite{paolacci2014inside,Palan_Schitter_2018,Buhrmester2018AnEO,thompson2020relational}, and thus the external validity of MTurk data has been investigated in various different research contexts~\cite{simons2012common,berinsky2014separating,goodman2013data,behrend2011viability,bartneck2015comparing} with varying results. Conclusions about the external validity of MTurk surveys about privacy and security have also been mixed: multiple studies~\cite{kang2014privacy,schnorf2014comparison} have found significant differences between an MTurk study and a U.S.-representative survey, with MTurk users reporting more concerns about privacy and information use and higher levels of social media use, while Redmiles et al.~\cite{redmiles2019well} found that that for participants under 50 years of age or with at least some college education, responses to questions regarding privacy and security were similar to the general population within these demographics, and that MTurk appeared to be more representative overall than a census-representative web panel. 

However, there have been noted concerns about demographic differences between the MTurk population and the over U.S. population. In particular, the MTurk population has been found to be younger and with higher levels of education than the overall U.S. population~\cite{paolacci2010running,ross2010crowdworkers,redmiles2019well,kang2014privacy}. Concerns have also arisen over the population on MTurk, particularly as highly active MTurk workers tend to complete many of the available tasks before others are able to, making the effective sample population on MTurk only 7000~\cite{peer2017beyond,stewart2015average}. Furthermore, while a study conducted in 2014 found that MTurk workers with over an 95\% approval rating provide high quality data and do not require attention checks~\cite{peer2014reputation}, more recent research has shown that data quality on MTurk has decreased dramatically to be less reliable than that on Prolific, even when quality filters (at least 95\% approval rating and 100 submitted tasks) were used~\cite{peer2021data}. 

\paragraph{Prolific}
Prolific was launched in 2014, and was primarily designed for use by researchers~\cite{prolific}. In the past few years, we have seen an increase in the use of Prolific as an alternative to conducting surveys on Amazon Mechanical Turk. Both the number of users and the number of researchers on the platform have increased dramatically in recent years~\cite{Palan_Schitter_2018, peer2021data}, and studies have shown that it is a viable alternative to MTurk~\cite{peer2017beyond}. 

Prolific mandates a minimum hourly payment for studies, and compensation may be adjusted by researchers if the survey takes longer than originally intended. Furthermore, users on Prolific have an option to return their submissions, indicating they no longer wish to participate or that they do not wish their data to be used, making it easy for participants to withdraw consent at any time.

Although a study in 2017 found Prolific to be slower in gathering responses than MTurk and CrowdFlower (another online survey site)~\cite{peer2017beyond}, we did not find such differences in our sample, perhaps due to the expansion of the Prolific worker pool over the last five years.

\begin{comment}
\paragraph{Other Platforms.}

CloudResearch, founded in 2016 by academic researchers primarily for recruiting user study participants, provides a toolkit that supports new tools and capabilities and runs on top of the MTurk platform. Among other features, CloudResearch screens MTurk workers in an effort to provide only high-quality (and human) study participants. Prior work has found that CloudResearch's MTurk Toolkit provides similar data quality to Prolific when quality filters (at least 95\% approval rating and 100 submitted tasks) were used~\cite{peer2021data}, and even superior to that of Prolific when CloudResearch's default recommended data quality filters were used, which samples the over 50,000 users in the US in the CloudResearch-approved group~\cite{litman2021reply}. 

Figure Eight, a human-in-the-loop platform previously known as CrowdFlower and now owned by Appen, is a crowdsourcing platform that focuses on data labeling and model evaluation for machine learning. Only limited work has looked at Figure Eight as a tool for recruiting human subjects; Peer et al.~\cite{peer2017beyond} found that Figure Eight workers failed more attention-check questions and had lower data quality compared to Prolific and MTurk. 

Neither CloudResearch nor Figure Eight have experienced significant adoption among the usable privacy and security community.
\end{comment}

\subsection{Data Quality and Attention Check Questions (ACQs)}

Some prior work has found that MTurk workers performed well on attention check questions~\cite{hauser2016attentive}, but other work found that MTurk workers were less attentive than convenience-sampled college students~\cite{goodman2013data}. Prior work comparing differing survey platforms in 2017 have found that almost half of MTurk and Prolific participants failed at least one attention check question, with MTurk users failing on average fewer attention checks than Prolific~\cite{peer2017beyond}. More recent work in 2021 saw Prolific users outperforming MTurk users on completing ACQs~\cite{peer2021data}. Excluding those based on passing attention checks had little effect for MTurk, and a small effect on Prolific~\cite{peer2017beyond}.

Prolific specifically allows for Instructional Manipulation Checks (IMCs), which are questions that ``explicitly instruct a participant to complete a task in a certain way'' such as clicking a specific answer~\cite{prolific}. IMCs and other attention checks have been shown to increase the reliability of data, and have become relatively widely used~\cite{oppenheimer2009instructional,harborth_evaluating_2021,hauser2015sa,peer2021data,chandler2019online}. However, IMCs might also influence participants to change interpretation and assessment of subsequent questions~\cite{hauser2015sa}.

Some research has also investigated comprehension, which involves checking that participants are able to understand instructions and explain them back to the researchers. This can be conducted in formats such as IMCs, or through textboxes asking users to summarize the instructions. However, these might not function exactly the same as attention checks, as prior work suggests those who fail attention checks may not be the same as those who do not comprehend instructions~\cite{berinsky2014separating}. Prior work has also found that Instructional Manipulation Checks making sure participants understood instructions improved data quality~\cite{crump2013evaluating}. Prolific users also tend to outperform MTurk users on comprehension checks, and there appears to be a positive correlation between correctly passing ACQs and comprehension questions~\cite{peer2021data}.

CAPTCHAs are commonly discussed as a mechanism for improving data quality by eliminating bots from a dataset, however prior work has found that bot accounts are able to reliably pass CAPTCHAs~\cite{moss2018after}.

\section{Methodology}

To evaluate the generalizability of online privacy and security surveys, we compared survey responses from four sources: (1) responses to a U.S. nationally-representative probabilistic sample, (2) people recruited through Prolific using their representative sample option, (3) people recruited through Prolific using their gender-balanced option, and (4) people recruited through Amazon Mechanical Turk (MTurk).

\subsection{Survey Questions}

To decide what questions to ask on our survey, we started by identifying categories of topics in privacy and security that have been the subject of recent user studies. We identified 28 papers published in the Proceedings on Privacy Enhancing Technologies Symposium (PoPETs) and the Symposium on Usable Privacy and Security (SOUPS) in 2021 that included user surveys. Two papers~\cite{danilova_code_2021,fulton_benefits_2021} exclusively surveyed specific technical populations (freelance developers and developers who have used Rust, respectively) about technical topics (security practices when developing code and experience with Rust), so we excluded them from our analysis. For the 26 papers that surveyed the public, we qualitatively coded the categories of questions asked in user surveys; we also determined what platform they used to recruit participants and how they handled attention check questions. 

Our qualitative coding identified five classes of questions that characterize the space of recent usable privacy and security surveys:
\begin{enumerate}
    \item \textbf{Behavioral.} Questions about what users do, would do, or have done in relation to technology, social media, and privacy and security tools. These questions refer to active behaviors undertaken by the user. For example, whether they use Twitter or whether they have recently decided not to use a service because of concerns about its data collection practices. 21 papers (80.8\%) included behavioral questions in their user survey~\cite{abrokwa_comparing_2021,bailey_i_2021,balash_examining_2021,ceci_concerned_2021,chandrasekaran_face-off_2021,cobb_i_2021,emami-naeini_understanding_2021,harborth_evaluating_2021,haring_never_2021,hasegawa_why_2021,kumar_designing_2021,lee_digital_2021,ray_warn_2021,smullen_managing_2021,story_awareness_2021,tahaei_deciding_2021,tang_defining_2021,tolsdorf_exploring_2021,wash_knowledge_2021,zhang_did_2021,zhang-kennedy_whether_2021}.
    \item \textbf{Experience.} Questions about whether or how often participants had experienced a particular type of event. These questions refer to actions or circumstances that occur to the respondent without active action on the part of that person. For example, how often they had experienced someone taking over their social media or email account without their permission or how often they were asked to agree to a privacy policy. 17 papers (65.4\%) included experience questions in their user survey~\cite{abrokwa_comparing_2021,bailey_i_2021,balash_examining_2021,cobb_i_2021,emami-naeini_understanding_2021,harborth_evaluating_2021,haring_never_2021,hasegawa_why_2021,kariryaa_understanding_2021,kaushik_how_2021,kumar_designing_2021,ray_warn_2021,smullen_managing_2021,story_awareness_2021,tahaei_deciding_2021,wash_knowledge_2021,zhang-kennedy_whether_2021}.
    \item \textbf{Knowledge.} Factual questions relating to privacy and security topics that test how much participants know about the topic. These questions have factually correct answers. For example, what it means if a website uses cookies or what a privacy policy is. 11 papers (42.3\%) included knowledge questions in their user survey~\cite{abrokwa_comparing_2021,bailey_i_2021,balash_examining_2021,bui_automated_2021,harborth_evaluating_2021,haring_never_2021,kariryaa_understanding_2021,kaushik_how_2021,story_awareness_2021,stransky_limited_2021,tang_defining_2021}.
    \item \textbf{Perceptions.} Opinion questions about user perceptions of and attitudes towards practices and behaviors. These questions---which focus on what respondents believe a principal would do or the reasons why they believe the respondent would do something---include questions about trust, comfort, and mental models. For example, how confident they were that a company would follow what the privacy policy says it will do or how comfortable they are with companies using their data to help develop new products. 19 papers (73.1\%) included perception questions in their user survey~\cite{abrokwa_comparing_2021,balash_examining_2021,ceci_concerned_2021,chandrasekaran_face-off_2021,emami-naeini_understanding_2021,gros_validity_2021,harborth_evaluating_2021,haring_never_2021,hasegawa_why_2021,kariryaa_understanding_2021,kaushik_how_2021,kumar_designing_2021,smullen_managing_2021,story_awareness_2021,stransky_limited_2021,tahaei_deciding_2021,tang_defining_2021,wash_knowledge_2021,zhang-kennedy_whether_2021}.
    \item \textbf{Beliefs.} Opinion questions about what security options or privacy rights people ought to have. Beliefs questions focus on what the respondent thinks should be true rather than asking about perceptions of the current world. For example, whether people should have the right to remove potentially embarrassing photos or criminal history from publicly-available search records. 9 papers (34.6\%) included belief questions in their user survey~\cite{abrokwa_comparing_2021,balash_examining_2021,cobb_i_2021,gros_validity_2021,haring_never_2021,kaushik_how_2021,kumar_designing_2021,smullen_managing_2021,zhang-kennedy_whether_2021}.
\end{enumerate}
For each of the five categories of questions, we selected 4-8 questions from a database of questions used in a past Pew Research Center survey~\cite{pew2019} (a total of 30 questions). Drawing our questions from this source had two key advantages: (1) Pew questions are extensively validated before being deployed and (2) responses from a large-scale ($n=4,272$), nationally-representative survey conducted by Pew in June 2019 are publically available~\cite{pew2019}, precluding the need to deploy our own nationally-representative panel survey. To enable intercomparison, our online surveys closely followed Pew's methods: the phrasing of the questions were the same, the set of possible responses were the same, the order of the questions were the same---with randomization of question order or answer choices matching the Pew questionnaire---and there were no forced responses. 

Since the Pew dataset includes demographic information for each participant, we also included basic demographic questions at the end of our survey. To facilitate comparisons with the Pew survey, we used demographic questions that matched the demographics released in the Pew dataset.\footnote{Note that these questions do not reflect current best-practices for asking about gender~\cite{spiel2019better} or race~\cite{us2016collection}. Nonetheless, we believed that matching the Pew phrasing was critical in order to enable direct comparisons with responses to the Pew survey.}

Finally, we identified three common techniques for excluding bots from online survey populations: reading-based attention check questions (i.e., questions that require participants to select a particular answer, also known as IMCs)~\cite{oppenheimer2009instructional, harborth_evaluating_2021}, free-response text questions (survey responses are rejected if the answer is nonsensical, irrelevant, or clearly copy-pasted from the Internet), and CAPTCHAs. To allow us to evaluate the effect of these techniques on external validity, we added two additional questions to our survey: one reading-based attention check question and one free-response text question. We also required half of our participants (randomly selected) to successfully complete a CAPTCHA in order to submit the survey. 

The full text of the survey can be found in Appendix~\ref{appendix:survey}.

\subsection{Datasets}
We use four datasets: (1) a probabilistic dataset from the Pew Research Center panel~\cite{pew2019}, (2) a representative sample from Prolific (accurate to the US Census on age, sex, and race), (3) a gender-balanced sample from Prolific, and (4) a sample from Amazon Mechanical Turk (MTurk). We compensated online study participants \$1.50 for completing the survey, which we estimated to take 6 minutes. This was approved by the Institutional Review Boards of the authors' institutions.

We additionally collected two filtered samples of underrepresented populations that do not appear in the Pew demographic categories: Indigenous people and transgender people. We deployed surveys on Prolific using the platform's prescreening filters to only allow participants in these demographics to take the study. Over a period of 8 days, we received responses from 79 Indigenous users on Prolific, and 197 transgender users.

\begin{enumerate}

\item \textbf{Pew American Trends Panel Wave 49.} This dataset ($n=4,272$) was collected by Ipsos Public Affairs between June 3-17, 2019 on behalf of the Pew Research Center~\cite{pew2019}. The weighted estimates for this sample are believed to accurate to $\pm 1.87$ percentage points of the US population aged 18 and over.
Pew Research Center typically makes survey data publicly available on their website two years after the data collection, so this dataset became publicly available in 2021.

Participants in this survey were a subset of Pew Research Center's American Trends Panel (ATP)~\cite{pewATP}, a panel of more than 10,000 U.S. adults recruited and maintained by the Pew Research Center using state-of-the-art techniques.\footnote{Prior to 2018, panel participants were recruited at the end of a large, national, landline and cellphone random-digit-dial survey that was conducted in both English and Spanish. After 2018, ATP has relied on address-based recruitment to avoid the response-bias that has developed in telephone-based recruiting. It supports non-Internet connected participants by providing tablets that enable those people to take surveys.} This subset of the panel was chosen to be generally representative of the broader U.S. population; as this was a probabilistic survey, the resulting data was weighted to balance demographics to match the U.S. population (to compensate for any biases due to sampling and non-response). 

Our analyses treat this dataset as the gold standard for U.S. responses to our survey questions.

\item \textbf{Prolific Representative Sample.} We sampled U.S. participants ($n=800$) using Prolific's representative sample feature. This sample is stratified on age, sex, and ethnicity based on the simplified U.S. census~\cite{prolific}. The median time to complete the survey was 6.1 minutes. %We henceforth call this sample the Representative sample.

\item \textbf{Prolific Gender-Balanced Sample.} We sampled U.S. participants ($n=800$) on Prolific, balanced on gender (50\% male and 50\% female). Prolific has been noted to have demographics that skew younger and more female \cite{charalambides2021we}: currently within the U.S. sample space, there are over twice as many women on the platform as men. This survey took participants a median time of 5.3 minutes to complete. %We refer to this sample as the Gender-Balanced Sample. 
No participants are in both the Prolific representative and gender-balanced samples.

\item \textbf{MTurk Sample.} We collected a sample ($n=800$) from Amazon Mechanical Turk, with participation restricted to people located in the U.S. who have completed over 50 HITs and have over 95\% approval rate. We chose these filters as they are common practice for studies of this type deployed on the MTurk platform and believed to produce higher quality data~\cite{peer2014reputation,redmiles2019well}. Participants took a median time of 5.2 minutes to complete the survey. %This sample is referred to as the MTurk sample.
\end{enumerate}

\subsection{Analysis}
We used chi-square proportion tests ($\chi^{2}$) to compare response distributions. For each question, we ran $\chi^{2}$ tests to compare the distribution of answers for each sample (Prolific representative, Prolific gender-Balanced, MTurk) pairwise against the Pew data. 
% the p-values resulting from the $\chi^{2}$ tests, 
We also used Total Variation Distance (TVD) to quantify the distance between answer distributions in our surveys and the Pew data. 
% to each question from our various online survey samples and response to the same question for the general U.S. population (represented by the Pew dataset in most of our analyses). 

\paragraph{Total Variation Distance.} Total Variation Distance (TVD), defined as $\text{TVD}(P,Q) = 1/2\cdot \Sigma_i(P_i,Q_i)$, is a standard metric for quantifying the distance between two distributions~\cite{gibbs2002choosing}. Intuitively, it corresponds to the fraction of respondents who answer differently between the two samples. A TVD of 0 indicates that two distributions are identical; as the distributions become increasingly disjoint the TVD approaches 1.

% This continuous measure of distance allows us to get a sense of how far apart the distributions are, rather than whether or not they are different, as the p-values from the $\chi^{2}$ tests give.
% Neither $\chi^{2}$ nor TVD 
%a TVD of 1 indicates that two distributions are 100\% different and 

To illustrate the concept of TVD, we show how to calculate the TVD between the distribution of responses to the first knowledge question for the Pew sample ($\textit{know1}_{\textit{Pew}}$) and responses for the Prolific gender-balanced sample ($\textit{know1}_{\textit{Bal}}$).  In the Pew survey, $.626$ of the respondents answered correctly, $.093$ incorrectly, and $.282$ with ``Not sure''; in our representative Prolific sample, the proportions for correct, incorrect, and not sure were $.866$, $.028$, and $.106$ respectively. %The numbers are found in Appendix~\ref{appendix:tables} by \textit{know1}. 
%Thus the TVD between the two samples is:
Therefore,
\begin{eqnarray*} &&\hspace{-26pt}TVD(\textit{know1}_{\textit{Pew}},\textit{know1}_{\textit{Bal}}) \\
&=& \frac{1}{2}(|.626-.866| + |.093-0.028| + |.282-.106|) \\
&=& .2405
\end{eqnarray*}
%$\frac{1}{2}\cdot(|0.626-0.897| + |0.093-0.038| + |0.282-0.065|) = 0.272$
This TVD of $.2405$ shows that approximately one quarter of the responses were distributed differently between the Pew sample and the Prolific representative sample.

To show how we use TVD, consider a comparison between the survey questions \texttt{know1} and \texttt{exp5}. In both cases, a $\chi^{2}$ test indicates a significant difference in answers between the Pew sample and the Gender-Balanced Prolific sample. To contextualize this result, we calculate TVD values for both  pairs of distributions. Using the same definition as above, we find that $TVD(\textit{exp5}_{\textit{Pew}},\textit{exp5}_{\textit{Bal}}) = .111$.  The lower TVD for \texttt{exp5} provides evidence that the balanced Prolific sample may be closer to the Pew sample for the experience question ($\text{TVD} = .111$) than for the knowledge question ($\text{TVD} = .2405$).

\vspace{10pt}
We chose to use these two measures ($\chi^{2}$ tests and TVD) as both have strengths and weaknesses in their ability to provide insights into the representativeness of these platforms' participants. $\chi^{2}$ tests with $p$-values provide a thresholded measure of sameness or difference, while TVD provides us with a limited but valuable continuous measure of distance. For example, when $\chi^{2}$ tests show that answer distributions are statistically distinct for two question categories, TVD augments this analysis by providing a method of estimating whether the non-representativeness of one question category may be of larger magnitude than the other.

As prior work has found that online surveys were representative for populations under 50 years old or with at least a college level education~\cite{redmiles2019well}, we further explored whether online survey platforms were more representative of certain demographic subsets in the general US population. In particular, we separately analyzed populations aged 18-29, 30-49, and 50+ from each of our samples against the corresponding demographic groups in the Pew dataset. We also conducted an analysis divided by education level, classified into one of three categories: high school graduate or less, some college, college graduate+.

Since the responses ranged from binary to multiple choice to Likert-scale, we did not attempt to code answers into binary variables. For most questions, we kept the response codings as  presented to the user. For knowledge questions---which had one correct answer, 3-5 incorrect answers, and a ``Not sure'' option---we coded answers into three categories: Correct, Incorrect, Not sure. In the studies, participants were able to skip any question. As no more than 2.5\% of any question had blank answers, we chose to impute the answers for non-response. Any skipped attention check was coded as a failed attention check. For knowledge questions, any skipped question was classified as ``Not sure''. For all other questions, non-response was classified into the most negative category (e.g. ``Not at all confident'', ``No, do not use this''). To validate this choice of imputation, we ran an analysis as well on imputation where non-response was classified as the most positive (e.g. ``Very confident'', ``Yes, use this'') and found that TVD remained $\pm 0.01$ both within each category and  overall.

To understand whether attention checks are effective in improving data quality, we ran an analysis wherein only responses of those who passed each of the three attention check questions were included. To determine efficacy of reading attention checks, we compared only those who passed the reading attention check (selected ``Strongly agree'') against the Pew data. For the textbox attention check which asked users to define ``digital privacy'' in their own words, a researcher from our team coded all responses into accept, reject, and copy-paste. ``Reject'' responses referred to answers that either were nonsensical, unrelated to the question, or merely repeated the words ``digital privacy''. ``Copy-paste'' indicated responses that were plausible definitions of ``digital privacy'', but appeared verbatim 5 times or more throughout the sample or contained large chunks of text that were copied verbatim from these phrases. Under this coding, some other phrases were indeed often repeated, but were accepted if they appeared less than 5 times. A second researcher resolved cases where it was uncertain whether a response should be rejected. We removed all responses either coded as ``Reject'' or ``Copy-paste'' when analyzing samples that are said to have passed the textbox attention check.
% As only the MTurk sample had significant proportions of users that failed the textbox attention check (less than 1\% of responses were rejected for each of the Prolific samples), we conducted this analysis on the MTurk sample alone. 
For the CAPTCHA analysis, for each of our samples, we ran analyses comparing those in the sample who saw and passed a CAPTCHA (around 50\% for each sample) against the Pew dataset.

We conducted demographic raking using the R \texttt{anesrake} package, weighted by age, sex, education, and race to see if it would improve the generalizability of our samples. We used proportions from the 2017 American Community Survey~\cite{ACS} to match the demographic weighting used for the Pew dataset.

We were further interested in whether underrepresented groups had significantly different responses than the general population. %We first examined, within the Pew dataset, whether responses from these groups were different from responses for the rest of the population. For example, as those over 65 are often underrepresented in online study platforms, we compare the responses from all those over 65 in the Pew dataset to all those under 65 to determine whether older people have different responses from the general population. If so, then this tells us that researchers ought not to generalize conclusions and findings to all groups. 
We compared demographic groups from our Prolific representative sample to the same group from the Pew sample to investigate whether demographic groups on Prolific are representative of their respective group in the broader population. With our filtered samples of rare demographic subpopulations (Indigenous and transgender people), we compared our filtered sample to the Prolific representative sample. %This allows us to minimize error, as should the Prolific population be substantially different on certain measures than the Pew population, differences in comparing these groups on Prolific to the general US population may not only be due to different experiences and preferences of the underrepresented groups, but also differences between Prolific and the general US population pool.

\begin{table}[t!]
\centering
\begin{tabular}{ll|cc|ccc}
  \hline
 \multirow{2}{*}{Dem}& \multicolumn{1}{c}{\multirow{2}{*}{Response}}& \multicolumn{2}{c}{Pew (\%)}
& \multicolumn{3}{c}{Online Samples (\%)} \\
\cline{3-4}\cline{5-7}
 & \multicolumn{1}{c}{} & \multicolumn{1}{c}{Raw} & \multicolumn{1}{c}{Wgt} & Repr & Bal & MT \\ 
  \hline
Age & 18-29 & 16 & 20 & 23 & 45 & 22 \\ 
   & 30-49 & 31 & 33 & 34 & 43 & 66 \\ 
   & 50-64 & 31 & 26 & 30 & 9 & 10 \\ 
   & 65+ & 23 & 20 & 12 & 2 & 2 \\ \hline
  Edu & HS or less & 35 & 38 & 13 & 13 & 8 \\ 
   & Some College & 28 & 31 & 34 & 34 & 26 \\ 
   & College grad+ & 38 & 30 & 53 & 52 & 66 \\ \hline
  Race & White & 78 & 74 & 75 & 75 & 82 \\ 
   & Black & 11 & 12 & 13 & 4 & 13 \\ 
   & Asian & 3 & 4 & 7 & 12 & 3 \\ 
   & Mixed & 4 & 5 & 4 & 7 & 2 \\ 
   & Other & 4 & 5 & 2 & 3 & 1 \\ \hline
  Sex & Male & 44 & 48 & 49 & 50 & 68 \\ 
   & Female & 56 & 52 & 51 & 50 & 32 \\ 
   \hline
\end{tabular}
\caption{Demographic characteristics of the four datasets. Since the Pew dataset was a probabilistic sample, the weighted dataset (Wgt) was analyzed. For the online samples---Prolific representative (Repr), Prolific gender-balanced (Bal) and MTurk (MT)---raw data was analyzed except where we explicitly state that raking was applied.} \label{table:demographics}
\end{table}

\subsection{Methodological Limitations}
\label{sec:limitations}
Due to the statistical constraints surrounding sample size and power, smaller sample sizes necessarily have less statistical power. Thus, for smaller samples (e.g., CAPTCHAs,  underrepresented groups), we expect to find fewer instances of \textit{statistical} significance ($p$-values$<0.05$), implying that these samples more closely match the Pew dataset. However, this does not mean differences do not exist, but rather that they might be too slight to detect at lower sample sizes. Thus, we examine TVDs in conjunction with $p$-values in order to obtain a clearer picture rather than simply defaulting to $p$-values.

\begin{figure*}[t!]

     \begin{subfigure}[b]{\columnwidth}
      \centering
       \resizebox{\textwidth}{!} {\input{figures/Figure_MTurk_Behavior.pgf}}
       %\vspace{-15pt}
      \caption{Behavior Questions. Mean TVD = .30} 
      %TVDs from left to right: behav1 = 0.10, behav2 = 0.10, behav3 = 0.31, behav4 = 0.53, behav5 = 0.48.
      \label{fig:results:mturk:behavior}
    \end{subfigure}
    \hfill
    \begin{subfigure}[b]{\columnwidth}
      \centering
       \resizebox{\textwidth}{!} {\input{figures/Figure_MTurk_Experiences.pgf}}
       \vspace{-24pt}
      \caption{Experiences Questions. Mean TVD = .09}
      %TVDs from left to right: exp1 = 0.11, exp2 = 0.05, exp3 = 0.09, exp4 = 0.08, exp5 = 0.10
      \label{fig:results:mturk:experiences}
    \end{subfigure}

    \smallskip  
    
    \begin{subfigure}[b]{\columnwidth}
      \centering
       \resizebox{\textwidth}{!} {\input{figures/Figure_MTurk_Knowledge.pgf}}
      \caption{Knowledge Questions. Mean TVD = .22}
      %TVDs from left to right (Row 1): know1 = 0.17, know2 = 0.16, know3 = 0.16, know4 = 0.29 \newline
      %Row 2: know5 = 0.19, know6 = 0.37, know7 = 0.15, know8 = 0.24
      \label{fig:results:mturk:knowledge}
    \end{subfigure}
    \hfill
    \begin{subfigure}[b]{\columnwidth}
      \begin{center}
      \captionsetup{justification=centering}
       \resizebox{\textwidth}{!} {\input{figures/Figure_MTurk_Perceptions.pgf}}
      \caption{Perception Questions. Mean TVD = .11}
      %TVDs from left to right (Row 1): percep1 = 0.11, percep2 = 0.12, percep3 = 0.16, percep4 = 0.14 \newline
      %Row 2: percep5 = 0.10, percep6 = 0.07, percep7 = 0.11, percep8 = 0.09
      \label{fig:results:mturk:perceptions}
      \end{center}
    \end{subfigure}
    
    \smallskip
    
    \begin{subfigure}[b]{\columnwidth}
      \centering
       \resizebox{\textwidth}{!} {\input{figures/Figure_MTurk_Beliefs.pgf}}
      \caption{Beliefs Questions. Mean TVD = .10}
      %TVDs from left to right: belief1 = 0.19, belief2 = 0.15, belief3 = 0.06, belief4 = 0.00
      \label{fig:results:mturk:beliefs}
    \end{subfigure}
    \hfill
    \begin{subfigure}[b]{\columnwidth}
      \centering
      \includegraphics[width = .3\textwidth] {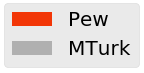}
      \caption{Legend. Brown is overlap of Pew and MTurk.}
    \end{subfigure}
    
      \caption{Distributions of responses to all questions for the Pew sample (weighted) and the MTurk sample (u50SC/Rak/freeAC). TVDs between the Pew sample and the MTurk sample are given in the captions.  %Best-case means that MTurk data is raked, filtered to those who passed the free-entry attention check, and limited to those reporting to be under 50 or having some college education. 
      }
      \label{fig:results:mturk}
\end{figure*}

% \begin{table}
%     \label{tab:results:mturk}
%     \begin{tabular}{lrrll}
%     \hline
%      Category     &   Raw TVD &   Best-case TVD & Raw \# p\ensuremath{>}0.5   & Best-case \# p\ensuremath{>}0.5   \\
%     \hline
%      Behavior     &      0.37 &            0.26 & 0/4           & 0/4                 \\
%      Beliefs      &      0.15 &            0.10 & 1/4           & 1/4                 \\
%      Experiences  &      0.27 &            0.09 & 0/5           & 0/5                 \\
%      Knowledge    &      0.30 &            0.22 & 0/8           & 0/8                 \\
%      Perceptions  &      0.30 &            0.11 & 0/8           & 0/8                 \\
%      Overall Mean &      0.28 &            0.16 & 1/30          & 1/30                \\
%     \hline
%     \end{tabular}

%     \caption{Total Variation Distance and number of questions which were statistically indistinguishable from Pew results ($p>0.05$) for each question category using raw results and best-case subset and quality controls. Lower TVD and higher $p>0.5$ counts indicate better representativeness (more closely matching answer distributions).These MTurk results are filtered to the best-case scenario for MTurk: under 50 or some college, with raking and free-response attention check filtering performed.}
% \end{table}

TVD is one of many measures that could be used to summarize our data and quantify distances between distributions. 
% Thus, our analysis primarily uses TVD not as an absolute measure but rather as a summary statistic to compare between different samples. 
A primary limitation of TVD is that it does not account for whether the underlying data is categorical or ordinal, and thus on Likert-scale style questions, treats participant answers which differ by a small ``amount'' (e.g., from 1 to 2) identically from those that differ by a large ``amount'' (e.g., from 1 to 5). Similarly, like $\chi^{2}$ tests, it cannot distinguish between different specific ways that categorical answer distributions differ. For example, TVDs for knowledge questions are large for both MTurk and Prolific representative ($.30, .23$), and $\chi^{2}$ tests find that responses to all 8 knowledge questions are significantly different from Pew responses for both, yet the \emph{direction} of these differences is opposite: MTurkers are more likely to be \emph{incorrect} in their knowledge while Prolific respondents are more likely to be \emph{correct}. Other options for contextualizing results from surveys are possible, such as visualizations and tables of raw answer proportions, which are provided in our figures and in Appendix~\ref{appendix:tables}. Qualitative work examining the reasons for and nuances of the differences we observe could  provide another avenue of understanding how and why results between probabilistic surveys and online platforms differ and what implications these differences have for the validity of insights provided by usable privacy and security studies. %In this work, we focus on the representativeness of samples from online platforms, though other contextual factors may also influence the validity of online surveys~\cite{xu2022contextualizing}.

We do not use corrections (e.g., Benjamini-Hochberg, Bonferroni) to analyze our results. These corrections control for Type I errors (false discovery rate) by limiting the number of erroneously statistically significant results ($p$-values $<0.05$). However, we are attempting to find results that \textit{do not} significantly differ between samples (i.e., $p>0.05$), so these corrections could in fact overstate our results.

\section{Results}

% Maybe better for attitudinal but not for experiences and knowledge

% Influence of pandemic pre/post samples?

% People who are online on these platforms are just a different 

% Results

We deployed three copies of our survey---Prolific representative sample, Prolific gender-balanced sample, and MTurk sample---in February 2022. We found that the Prolific representative survey took significantly longer to run; it took 49 hours to complete, compared to 2.5 hours for the Prolific gender-balanced survey and 2 hours for the MTurk survey.\footnote{However, we note that since a significant percentage of our MTurk responses (39.1\%) failed the free-response attention check, rejecting those responses and re-releasing those HITs would significantly increase the total deployment time; since less than 1\% of responses failed that check for either Prolific sample, no extra time or effort would be required for those surveys.} The Prolific representative survey also cost significantly more to deploy: collecting that sample cost \$2,784, compared to \$1,600.00 for the Prolific gender-balanced sample and \$1,682 for the MTurk sample. We then analyzed the responses to evaluate the external validity and data quality of the resulting samples. A summary of the demographics for each of the samples is provided in Table~\ref{table:demographics}, and complete results are summarized in Appendix~\ref{appendix:tables}.

\subsection{The External Validity of MTurk}
\label{sec:results:mturk}
    
Our MTurk sample was heavily weighted toward younger participants (703/800 participants were under 50) and those with higher education (737/800 participants had at least some college education); this finding replicates prior work~\cite{redmiles2019well, kang2014privacy}. However, while Redmiles et. al.~\cite{redmiles2019well} found that for the well-represented demographics (people under 50 or with at least some college) MTurker responses to privacy and security questions (about behavior, experiences, and knowledge) had high external validity, we were unable to replicate that result.

When we analyzed the raw MTurk sample, we found that responses collected through MTurk were extremely different from Pew (Table~\ref{tab:results:mturk}). We found statistically significant differences in responses for 29 of the 30 questions, and the overall average Total Variation Distance (TVD) was .29 (intuitively indicating that more than a quarter of the sample answered differently). We attempted to replicate prior work by also analyzing the sample that contained only people with under 50 or some college education and applied raking (i.e., demographic weighting). However, we still found statistically significant differences in responses for 29 questions, and the average TVD dropped only slightly (to .28). 

Unlike the earlier work, we found that both raking and filtering out responses that failed a free-response text attention check question had significant effects on data quality. Combining these data quality measures with the demographic restrictions from prior work---i.e., restricting to people under 50 or with some college who passed the text attention check and applying filtering, denoted u50SC/Rak/freeAC in Figure~\ref{fig:results:mturk} and Table~\ref{tab:results:mturk}---produced the best-case results for the MTurk sample. 

In this best-case scenario, 29 of the questions still had significantly different responses, but the average TVD went down to .17. In particular, responses about experiences, perceptions, and beliefs are somewhat generalizable: TVDs dropped to around $.10$, although $\chi^2$ tests still showed that responses for most questions were still significantly different from Pew. However, knowledge questions were still very different: MTurkers were much more confident---and incorrect---on knowledge questions even after data quality measures were applied. Behavior questions---which focused on social media use---also remained very different: MTurkers were more likely to use Facebook (63\% vs. 53\%), Instagram (74\% vs. 44\%), Twitter (79\% vs. 26\%), and other social networks (82\% vs 34\%). Complete results for this best-case scenario are depicted in Figure~\ref{fig:results:mturk} and measures of external validity are summarized in Table~\ref{tab:results:mturk}.

\begin{table}
    \begin{center}
    \begin{tabular}{l|cc|cc}
    \hline
     \multirow{2}{*}{Category}    &  \multicolumn{2}{c|}{Raw Sample}  & \multicolumn{2}{c}{u50SC/Rak/freeAC}\\\cline{2-5}
     & TVD & $p<0.05$   & TVD & $p<0.05$   \\
    \hline
     Behavior     &      0.41 & 5/5      &         0.30        & 5/5                 \\
     Experiences  &      0.27 & 5/5     &         0.09        & 5/5                 \\
     Knowledge    &      0.30  & 8/8   &         0.22        & 8/8                 \\
 Perceptions  &      0.30    & 8/8    &         0.11        & 8/8                 \\ 
     Beliefs      &      0.15 & 3/4   &         0.10        & 3/4                 \\
    \hline
     Overall  &      0.29 & 29/30   &         0.17        & 29/30                \\

    \end{tabular}
\end{center}
    \caption{Measures of external validity for the MTurk sample. TVD indicates distance from the (weighted) probabilistic Pew sample. $p<.05$ shows the fraction of questions in each category for which the responses were statistically significantly different from the Pew sample. For both, lower is better. %External validity measures are shown for both the raw sample and for the MTurk dataset filtered for only participants under 50 or with some college education who passed the free-response ACQ after raking was applied (u50SC/Rak/freeACQ).
    }
    \label{tab:results:mturk}
\end{table}

%    We note that the general trends of non-representativeness for MTurkers which we have described above are largely shared by Prolific respondents (see Section~\ref{sec:results:prolific}), with one exception for \textbf{Knowledge} questions. Both Prolific and MTurk respondents are more confident (less likely to give ``Not Sure'' as a response), but MTurkers are more often incorrect while Prolific respondents more often correct than the general population. Aside from those Knowledge results, the trends of Prolific and MTurk's differences from Pew data are generally shaped similarly, but the magnitude of differences for MTurk are much larger.

\subsection{The External Validity of Prolific} 
    \label{sec:results:prolific}

    Like the MTurk sample, the Prolific gender-balanced sample was heavily weighted towards younger participants and those with higher education; the age skew was more extreme and the education skew less. That sample also included significantly fewer Black participants and more Asian and mixed race participants. The Prolific representative sample was representative of the overall U.S. population for age and race, but showed the same skew towards higher education.

\begin{figure*}[t!]
     \begin{subfigure}[b]{\columnwidth}
     \begin{center}
      \resizebox{\textwidth}{!} {\input{figures/Figure_Prolific_Behavior.pgf}}
      %\vspace{0pt}
      \caption{Behavior Questions. Mean TVD (Repr, Bal) = .22, .27}
      \label{fig:results:prolific:behavior}
    \end{center}
    \end{subfigure}
    \hfill
    \begin{subfigure}[b]{\columnwidth}
     \begin{center}
      \resizebox{\textwidth}{!} {\input{figures/Figure_Prolific_Experiences.pgf}}
      \vspace{-24pt}
      \caption{Experiences Questions. Mean TVD (Repr, Bal) = .07, .06}
      \label{fig:results:prolific:experiences}
    \end{center}
    \end{subfigure}
    
    \smallskip
    
    \begin{subfigure}[b]{\columnwidth}
      \begin{center}
      \resizebox{\textwidth}{!} {\input{figures/Figure_Prolific_Knowledge.pgf}}
      \caption{Knowledge Questions. Mean TVD (Repr, Bal) = .23, .24}
      \label{fig:results:prolific:knowledge}
      \end{center}
    \end{subfigure}
    \hfill 
    \begin{subfigure}[b]{\columnwidth}
      \begin{center}
      \captionsetup{justification=centering}
      \resizebox{\textwidth}{!} {\input{figures/Figure_Prolific_Perceptions.pgf}}
      \caption{Perception Questions. Mean TVD (Repr, Bal) = .05, .05}
      \label{fig:results:prolific:perceptions}
      \end{center}
    \end{subfigure}
    
    \smallskip
    
    \begin{subfigure}[b]{\columnwidth}
      \begin{center}
      \resizebox{\textwidth}{!} {\input{figures/Figure_Prolific_Beliefs.pgf}}
      \caption{Beliefs Questions. Mean TVD (Repr, Bal) = .07, .08}
      \label{fig:results:prolific:beliefs}
      \end{center}
    \end{subfigure}
    \hfill 
    \begin{subfigure}[b]{\columnwidth}
      \centering
      \includegraphics[width = .3\textwidth] {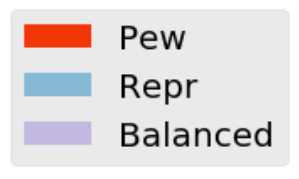}
      \caption{Legend. Dark purple is overlap of all 3.}
    \end{subfigure}
    
    \caption{Distributions of responses to all questions for the Pew sample (weighted), Prolific representative sample (raw), and Prolific gender-balanced sample (raw). %Raw Prolific data is used, without the application of any data quality measures or demographic limitations. 
    TVDs between the Pew sample and the Prolific samples are given in the captions. %The full text of all questions is given in Appendix~\ref{appendix:survey}.
    }
    \label{fig:results:prolific}
\end{figure*}

Overall, we found that both Prolific samples generalize better than the MTurk sample and that free-response attention checks were no longer critical for data quality. However, the external validity of the samples varied significantly depending on the type of question. Our results are shown in Figure~\ref{fig:results:prolific}, and measures of external validity are summarized in Table~\ref{tab:resultsprolific-infigures}.
    
    \textbf{Behavior.} Although responses about behavior from the Prolific representative sample were slightly more generalizable in terms of TVD (and both were more generalizable than the MTurk responses), neither of our Prolific samples demonstrated high external validity for behavior questions.
    Prolific participants were similarly likely to use Facebook compared to Pew participants but differed on other reported behavior, with the fraction of Prolific participants reporting that they use Instagram, Twitter, and ``Other Social Media Sites'' being 25\%--54\% higher compared to the Pew sample.

\begin{table}
    \label{tab:results:prolific}
    \begin{tabular}{ll|cc|cc}
\hline
\multirow{2}{*}{Samp.}    & \multirow{2}{*}{Cat.}    
&  \multicolumn{2}{c|}{Raw Sample}  & \multicolumn{2}{c}{u50SC/Rak/freeAC}\\\cline{3-6}
    & & TVD & $p<0.05$   & TVD & $p<0.05$   \\
    \hline
 Rep. & Behav.    &      0.22 & 4/5          &          0.19 & 3/5              \\
 Rep. & Exp. &      0.07 & 5/5          &          0.06 & 5/5              \\
 Rep. & Know.   &      0.23 & 8/8          &          0.17 & 8/8              \\
 Rep. & Percep. &      0.05 & 3/8          &          0.06 & 4/8              \\
 Rep. & Beliefs     &      0.07 & 3/4          &          0.06 & 2/4              \\ \hline
 Rep. & Overall & 0.13  & 23/30 & 0.11 & 22/30 \\
 \hline\hline
 Bal.       & Behav.    &      0.27 & 3/5          &          0.24 & 4/5              \\
 Bal.       & Exp. &      0.06 & 4/5          &          0.05 & 4/5              \\
 Bal.       & Know.   &      0.24 & 8/8          &          0.16 & 7/8              \\
 Bal.      & Percep. &      0.05 & 4/8          &          0.07 & 7/8              \\
 Bal.       & Beliefs     &      0.08 & 2/4          &          0.06 & 3/4              \\
\hline
Bal. & Overall & 0.14 & 21/30 & 0.12 & 25/30 \\
\end{tabular}

    \caption{Measures of external validity for the Prolific samples. For both, lower is better. See Table~\ref{tab:results:mturk} for more details.}\label{tab:resultsprolific-infigures}
\end{table}

    \textbf{Experiences.} Overall, both of our Prolific samples generalized well for questions about prior experiences.  While most of those questions showed statistically significant differences, the magnitude of those difference was quite small ($\text{TVD}=.07$ for the representative sample and $.06$ for the gender-balanced sample).
        
    \textbf{Knowledge.} Knowledge questions did not generalize well for either of our Prolific samples. Prolific respondents were more likely to provide \textit{correct} answers and less likely to answer ``Not sure'', suggesting that Prolific users are significantly more knowledgeable about privacy and security matters than the overall U.S. population.

    \textbf{Perceptions.} Both Prolific samples had relatively high external validity for questions about perceptions of privacy and security. The Prolific representative sample had statistically different responses compared to the Pew sample for only 3/8 questions (4/8 for the gender-balanced sample), and TVDs between each sample and the Pew sample were small (about .05). %The largest difference occurs in p5, for which, compared to Pew, 12\%/14\% more participants in each Prolific sample said they were ``Not Confident At All'' that companies would be held accountable by the government for misuse or compromise of personal data. Unlike MTurk, which shows a clear trends greater confidence and comfort (see Figure~\ref{fig:results:mturk:perceptions}, whose histograms skew to the left toward Very Comfortable/Somewhat Comfortable for MTurk), no obvious trend is present in Prolific respondents' confidence and comfort (see Figure~\ref{fig:results:prolific:perceptions}).
    
    \textbf{Beliefs.} Both Prolific samples also had high external validity for questions about beliefs about privacy and security. While some of the questions were statistically distinguishable, the TVDs were low suggesting that the effect size was small.
    
We also analyzed the Prolific samples using the best-case data quality measures from our MTurk analysis---restriction to people under 50 or with some college who passed the text attention check and applying filtering, denoted u50SC/Rak/freeAC in Table~\ref{tab:resultsprolific-infigures}). Overall, TVDs decreased slightly. However, unlike for the MTurk sample, this analysis did not dramatically improve the generalizability of the Prolific samples.

    \subsection{Data Quality Measures}        \label{sec:results:prolific:dataQuality} 
    We  evaluated four data quality measures: reading-based attention checks, free-response text attention checks, CAPTCHAs, and raking. For the MTurk sample, we found that a free-response attention check (which 39.1\% of responses failed) and raking both significantly improved data quality for the MTurk sample. Despite the data quality issues with MTurk, neither reading-based attention checks nor CAPTCHAs (which no respondents failed) significantly improved data quality. Although Prolific respondents did slightly less well than MTurkers on our reading attention check question (7.75\%-8.25\% failed), none of our data quality measures significantly improved data quality for the Prolific samples.

\subsection{Beyond the ``Average'' User}

While the standard metric of external validity is the extent to which results generalize to the overall population, overall generalizability does not necessarily imply that results are valid across all subgroups. We therefore also examine the question of how well our results generalize for various demographics subpopulations. We apply two analysis techniques: (1) we compare demographics slices from our online samples to the corresponding demographic slices of the Pew sample, an approach that parallels the investigations of MTurk generalizability by Redmiles et al.~\cite{redmiles2019well} and (2) we compare demographic slices from rarer, traditionally understudied subpopulations to the overall population.

\subsubsection{Prolific vs. Pew Demographic Subpopulations}

Since Pew is our gold standard, we can perform this analysis only for demographic variables reported by Pew, and only for values of those variables which occur sufficiently frequently  in the population to enable meaningful comparison. Based on these limitations, we choose to analyze two racial subpopulations (Black and Asian American), educational attainment, and age. Table~\ref{table:demographics} presents the numbers of people in each of these categories. Because the sample sizes are inherently smaller for these subpopulations than for the overall population, we focus our analysis exclusively on distance between the distribution of responses provided in the online surveys and the (weighted) distribution of responses to the Pew survey instead of considering $p$-values or the number of questions with statistically different responses.

Overall, we found that the Prolific representative sample tends to be the best of all three collected samples for each of these demographic brackets (although the Prolific gender-balanced sample is often nearly as good) and that the Prolific samples generalize better for younger and for more highly-educated subpopulations. 
    
\paragraph{Race.} Overall, the Prolific representative sample is almost as representative of the Black and Asian subpopulations as it is for the overall population. Average TVDs measuring the representativeness Black and Asian subpopulations are about .01--.02 higher on average than TVDs for the full dataset. %For the subpopulations of people over 65 and people with a high school degree or less education, TVDs are 0.03--0.04 higher, slightly worse overall than for our racial groups.

\paragraph{Age.} For Prolific, we found that as ages increase, the samples become less generalizable. For people age 18-29, both Prolific samples are fairly representative of the general U.S. population, with particular improvement for knowledge questions ($\text{TVD} = .15$) and behavior questions ($\text{TVD} = .16$). Within  both 18-29 and 30-40 age brackets, the Prolific samples actually generalize to the Pew dataset better than comparing the full datasets. By contrast, generalizability for people over 50 is worse, particularly for knowledge questions ($\text{TVD: Repr} = .28, \text{Bal} = .32$) and behavior questions as older Prolific users demonstrate significantly more knowledge of privacy and security and significantly higher levels of technology use. For our MTurk sample, both the 18-29 and the over 50 subpopulations were more generalizable than the full dataset, although they still had lower data quality than the corresponding slices of the Prolific samples. %Within our age brackets, the Prolific samples generalize better to the Pew dataset for ages 18-29 than overall when comparing the full dataset to the full dataset, similarly to the overall for ages 30-49, and worse for those ages 50+. This suggests a trend that as ages increase, Prolific samples become less generalizable.

\paragraph{Education.} For our Prolific samples, we found that as education increases the samples become more generalizable. For respondents with a high school education or less, Prolific samples are less generalizable for this demographic slice than for the overall population, with participants reporting particularly higher levels of technology use. For respondents who are college graduates, both Prolific samples are reasonably representative of the overall population of U.S. college graduates ($\text{TVD: Repr} = .10, \text{Bal} = .12$), with more representative responses to knowledge questions ($\text{TVD: Repr} = .13, \text{Bal} = .14$). Conversely, the generalizability of the MTurk sample for people with high school education or less did not decrease (although the data quality remained worse than the Prolific samples) while data quality does decrease slightly for the subpopulation with Bachelor's degrees ($\text{TVD} = .30$).

\subsubsection{Rare Demographic Subpopulations}

Finally we identified two populations with relatively low representation on Prolific---Indigenous people and transgender people---and explored (1) how effective Prolific's filters are at producing large samples of rare (and frequently understudied) subpopulations and (2) to what extent generalizable results for the overall population apply to these subpopulations. 

\begin{comment} % this info is in the text, don't need a table
        \begin{table}
            \label{tab:results:differentStats}
            \begin{tabular}{l|lrlrl}
                \hline
%                Within Pew & n & & & \\
%                \hline
%                Black & 458 & & & \\ 
%                Asian & 125 & & & \\
%                H.S. Graduate or Less & 1483 & & & \\
%                Over 65 & 977 & & & \\
%                \hline
                Filtered Samples & n & Days & n/Day & Avail. \\
                \hline
                Indigenous People     & 79    & 8     & 10 & 294 \\
                Transgender People    & 197   & 8     & 25 & 1231 \\
                \hline
            \end{tabular}
            \caption{Statistics for our analyses of the differences of specific populations. n indicates sample size. For filtered samples, our survey was active for ``Days'', and collected an average of ``n/day'' responses per day. Prolific's filtering tool indicated that ``Avail.'' respondents with the given demographic characteristic (transgender or Indigenous) had been active on Prolific in the past 90 days.}
        \end{table} 
\end{comment}

        \paragraph{Indigenous People}        
        Our filtered sample of Indigenous people ran for 8 days and obtained 79 responses during that time, an average of about 10 people per day. For context, at the time that we launched this filtered sample, Prolific reported that there were 294 Indigenous respondents who had been active in the past 90 days.
        
        Comparing the distributions of responses of these 79 respondents to our full Prolific representative sample, we found that variations were relatively small, with Indigenous people on Prolific being generally somewhat similar to other people completing surveys on Prolific. Comparable to the difference between our Prolific representative sample and Pew on the most representative question categories, mean TVDs comparing Indigenous respondents and the general Prolific population were under $.10$ for all question categories. 
        
        We emphasize that given the small size of this sample, we are unable to make conclusive statements about trends among Indigenous people on Prolific. Generally speaking, our data supports the claim that Indigenous people are more similar than different to other Prolific users, with TVDs between .04 and .05 for 3 question categories (Experiences, Perceptions, Beliefs). In terms of behaviors, they are more likely to use all social networks, including especially Facebook ($\text{TVD} = .16$) and other social networks ($\text{TVD} = .12$). Indigenous people in our sample appear to be slightly more likely to answer ``Not Sure'' to knowledge questions. No other clear trends emerge in how Indigenous respondents on Prolific are different from other respondents on Prolific in terms of experiences, perceptions, or beliefs.
        
        \paragraph{Transgender People}
        Our filtered sample of transgender people ran for 8 days and obtained 197 responses during that time, an average of about 25 people per day. At the time that we launched this filtered sample, Prolific reported that there were 1231 transgender respondents who had been active in the past 90 days.
        
        Comparing the distributions of responses of these 197 respondents to our full Prolific representative sample, we find that variations are small to moderate, with transgender people on Prolific being generally somewhat similar to other people completing surveys on Prolific. Mean TVDs comparing transgender Prolific respondents to the overall Prolific representative sample were under .12 for all question categories.
        
        Although the low sample size precludes definitive findings, our data for this subpopulation provides preliminary evidence of some potential interesting trends.  In terms of behavior, transgender people were less likely to use Facebook, and more likely to use Instagram, Twitter, and other social networks, than the Prolific Representative population. Transgender people were also more knowledgeable than Prolific participants overall, answering 6/8 knowledge questions correctly more often. Notably, transgender people were particularly more likely to understand how private browsing works, with a very large TVD of .26 distinguishing them as much more likely to answer \texttt{know8} correctly and much less likely to answer incorrectly or with ``Not Sure''. This result might be due to the need for transgender people to use private browsing mode to protect themselves and their browsing habits from local adversaries, such as family, while seeking community, gathering information, and engaging in activism online~\cite{lerner2020privacy}.
        
        Transgender people were also consistently more likely (TVDs between .07 and .13) to select ``Not Confident At All'' in response to perception questions that asked about confidence that companies will follow their privacy policies, promptly notify about data breaches, publicly admit mistakes that lead to privacy breaches, use personal information in appropriate ways, and be held accountable by the government for privacy missteps (\texttt{percep1--percep5}). Finally, they were slightly more likely to believe that people should have the  right to have various personal information removed from public search results, with a particularly large ($\text{TVD} = .16$) increase compared to the Prolific representative sample in the likelihood that transgender people would say that people should be able to have ``Negative media coverage'' about themselves removed from public search results. We hypothesize that this might emerge from the likelihood of transgender people to experience media coverage about them as negative, for example if articles misgender them or include out-of-date personal details such as deadnames.
\section{Discussion}

While our results quantify the external validity of online surveys about privacy and security, they also provide insight into best practices for  conducting online studies in this space. 

\vspace{12pt}
\noindent\textit{\textbf{Recommendation 1:} We recommend preferring Prolific to MTurk when recruiting participants for privacy and security surveys.}\\
    
Overall, we found major degradation of MTurk data quality and external validity since 2017~\cite{redmiles2019well}.
% Stringent open textbox attention checks (which removed 39\% of responses) did improve data quality significantly (see Section~\ref{sec:results:mturk}), 
If MTurk samples are used, applying the data quality measures studied in this paper---including demographic raking and a stringent open textbox attention check---is critical to enhance external validity. However, even when applying these data quality measures to MTurk data, Prolific gender-balanced samples provide better validity and their use is recommended at the current time. It is important to remember, however, that both Prolific and MTurk samples better represent younger and more educated populations. Additionally, online samples appear to be differently representative for different types of questions, as we discuss below in the Recommendation 3.

Future work should examine the validity of samples from other platforms, which could be comparable to or better than Prolific. For example, we note that CloudResearch, which uses the MTurk population, has been found to provide similar data quality to Prolific when the default data quality filters are applied~\cite{litman2021reply}. %In our literature review, %review of 26 recent papers in SOUPS and PoPETs that used online samples, 
%we did not find any papers that used CloudResearch, and thus did not include it in our analyses for this paper. However, we plan to consider CloudResearch in future work. 
Although our literature review did not find any papers that used CloudResearch, it it provides an alternative platform for recruiting participants in the future.
Future work should also continue to monitor the external validity of MTurk and Prolific, as population demographics and data quality may continue to change over time.

\vspace{12pt}
\noindent\textit{\textbf{Recommendation 2:} Determinations about whether to use Prolific's representative sample feature can make trade-offs between generalizability and logistical constraints without significantly impacting data quality for most studies.} \\ 

%\subsection{Our Best Practices}
We find that Prolific's representative sample feature produces data that most closely matches the results from the nationally representative sample from the Pew dataset. However, the representative sample takes much longer (49 hours vs. 2.5 hours for 800 responses) and is significantly more expensive (\$2784 vs. \$1600) to deploy than collecting a gender-balanced sample of the same size from Prolific. In most cases, a Prolific  gender-balanced sample performs nearly as well as a representative sample, with less than .02 difference in average TVD across all question categories. The largest gains for representative over gender-balanced were for behavior questions, for which neither was representative.  All other question categories had very small differences ($\text{TVD} < .01$) between representative and gender-balanced samples.
    
    %Add something here about trends for age and education. Over 50 not a hard cutoff, simply more representative if younger (18-29 better than 30-49)
    %HS or less not a hard cutoff, but rather less representative as education attainment decreases

%\subsubsection{Online Samples are Not Representative of Knowledge and Social Media Use}
\vspace{12pt}
\noindent\textit{\textbf{Recommendation 3:} Be cautious when drawing conclusions from online studies about privacy and security knowledge or social media use, as these results might not be representative of the overall U.S. population.}\\

None of our online samples were representative of the overall population for knowledge questions---which posed factual questions about privacy and security topics---or behavior questions---which were dominated by questions about rates of social media use. We recommend that researchers take care in designing studies and interpreting data which depend on these properties of respondents. Similar to prior work, we do find that the younger and more highly educated the population, the more Prolific is representative, particularly for knowledge questions, which drop from TVDs of .28 for those over 50 to .15 for those 18-29, and from .27 for those with high school degrees or less education to .13 for those with college degrees. Even these TVDs are quite high, however, indicating that 15-13\% of responses are different than they would be for that actual age or education range in an census-representative sample, and so we still urge caution in relying on such data.

Our results show that participants recruited through MTurker and Prolific are more confident about privacy and security knowledge compared to the overall U.S. population, with fewer respondents answering ``Not sure'' to most questions. We observe that this is a similar phenomenon to past results which found that MTurk workers are more certain about what information is available about them online~\cite{kang2014privacy}. This confidence gap also raises questions about the generalizability of non-survey studies that recruit participants through these online platforms, since prior work has found that confidence is a better predictor of security behaviors than actual knowledge~\cite{sawaya2017self}.

One factor that might have contributed to the drastically higher reported use of social media in our online samples is response bias from participants who worry that they may be excluded from a survey (and thus not be paid) if they don't use certain products, leading them inaccurately claim that they use social networks which they actually do not. Another possibility is that the population on these platforms have different behaviors regarding social media than the general U.S. population, leading to higher adoption and use of social media platforms. Indeed, prior work on MTurk has also found that U.S. MTurk workers have higher reported social media use than the general US population~\cite{kang2014privacy}, which is supported by our findings in our Prolific and MTurk samples. 

While it is possible that some of the difference in responses to behavior questions might have been due to actual differences in social media use between 2019 and 2022, a 2021 survey~\cite{auxier2021social} conducted by Pew about a year into the pandemic shows that social media adoption has not drastically increased since mid-2019, when the American Trends Panel Wave 49 was conducted. That survey found that in 2021, 69\% of Americans used Facebook, 40\% used Instagram, and 23\% used Twitter, numbers which are very closely compatible with the 71\%, 38\%, and 23\% found in 2019. This suggests that the  higher usage numbers we find in our online samples are genuine symptoms that users of online crowdsourcing platforms are not representative of the overall population in terms of their social media use.
    
\vspace{12pt}
\noindent\textit{\textbf{Recommendation 4:} Attention Check Questions and CAPTCHAs are not recommended for online surveys conducted on Prolific.} \\

We do not recommend reading attention checks (Instructional Manipulation Checks), textbox attention checks, and CAPTCHAs when collecting survey responses on Prolific. Our reading attention check was failed by 66/800 (representative sample) and 62/800 (gender-balanced sample) participants, but data quality was not improved by analyzing the data with these responses removed (see Section~\ref{sec:results:prolific:dataQuality}). Prolific users almost never fail textbox attention checks (7/1600 failures) or CAPTCHAs (0/1600 failures).  Based on our results, using such attention check questions lengthens surveys unnecessarily. Using IMCs might also change participants interpretation of subsequent questions~\cite{hauser2015sa}.
    
\vspace{12pt}
\noindent\textit{\textbf{Recommendation 5:} Raking is not currently necessary when analyzing the results of online privacy and security studies.} \\

Raking is often used in survey methods that intend to be representative of the general population since perfect response rates from demographic groups cannot be ensured by any sampling approach.  Although we found raking had little effect on the representativeness of either of our Prolific samples (Section~\ref{sec:results:prolific:dataQuality}), studies in other disciplines have seen success in using raking for MTurk survey data to better generalize to the US population~\cite{simons2012common}, and we would recommend researchers consider it. However, we note that raking also requires decisions on which demographic variables to weight on and might differ for different questions and fields of study.

\vspace{12pt}%\pagebreak
\noindent\textit{\textbf{Recommendation 6:} Special care should be taken to include a diverse population of study participants, particularly for demographics that are rare or underrepresented  on crowdsourcing platforms.} \\

Prior work has noted that online platforms tend to be younger, more highly educated, and more white than the general U.S. population~\cite{redmiles2019well}. This demographic imbalance could then lead to a fallacy of the ``average'' user on such platforms not being at all representative of the general population. Groups that do not make up the majority might also have significantly different preferences than the ``general population''. For example, participants from racial minorities were more unsure about their security knowledge than the general population, and transgender people had lower confidence in companies taking responsible action regarding privacy issues than the general Prolific population. Therefore, we encourage researchers to consider specifically sampling underrepresented populations to understand possibly divergent privacy and security perceptions and backgrounds to avoid over-general interventions and claims that could contribute to further marginalizing already marginalized populations.
    
\paragraph{Study Limitations.}
As we limited our studies to participants located in the United States, we cannot make claims as to whether Prolific is similarly representative of other jurisdictions or of the overall global population. Indeed, prior work has also found differences between privacy attitudes between MTurk workers located in the U.S. and in India~\cite{kang2014privacy}. Additionally, while the Pew dataset was weighted on myriad strata of demographics to best represent the adult U.S. population, we still recognize that it is not perfectly representative of the general US population. As with all surveys, there might be non-response bias, even with the most carefully selected probabilistic sample. In other words, just as those who choose to take surveys on online platforms differ from the general population in terms of tech familiarity and use, so too might probabilistic studies vary from exact national preferences. Therefore, though we use Pew as our gold standard, we cannot guarantee that it is a perfect representation of the preferences of all Americans.

\section{Conclusion}

Online crowdsourced samples are an important source of data for usable privacy and security survey research today. %, and they enable large samples to inform research on a variety of critical topics. 
Understanding the external validity of these samples is critical to ensuring that the results from such research generalize and can appropriately guide individuals, technologists, lawmakers, and regulators. 
Our work evaluates the external validity  two popular crowdsourcing sites---MTurk and Prolific---and provides recommendations about best practices for conducting security and privacy surveys on these platforms. 

\section*{Acknowledgments}
We are grateful to Cassandra Pattanayak and Elissa Redmiles for their advice and support on this work. This work was supported by the National Science Foundation (CNS Award \#1948344) and Wellesley College.

%-------------------------------------------------------------------------------
\bibliographystyle{plain}
\bibliography{usenix2022_SOUPS.bib}

\appendix
\section{Survey Questions}\label{appendix:survey}

This appendix contains the list of questions asked during our user study. These questions are taken from the Pew American Trends Panel run between June 3-17, 2019. Each question could be skipped by the user.

%The survey questions are as follows:\\

\begin{enumerate}
    \item Do you use any of the following social media sites?\\
    \indent \textit{The order of the first three of the following questions are randomized}
    \begin{enumerate}
        \item Facebook \textit{[behav2]}
            \begin{itemize}
                \item Yes, use this / No, do not use this
            \end{itemize}
        \item Instagram \textit{[behav3]}
            \begin{itemize}
                \item Yes, use this / No, do not use this
            \end{itemize}
        \item Twitter \textit{[behav4]}
            \begin{itemize}
                \item Yes, use this / No, do not use this
            \end{itemize}
        \item Any other social media sites \textit{[behav5]}
            \begin{itemize}
                \item Yes, use this / No, do not use this
            \end{itemize}
    \end{enumerate}
    \item In your own words, what does “digital privacy” mean to you?\\
    \indent \textit{Participants are given a textbox to type in.}
    \item How often are you asked to agree to the terms and conditions of a company’s privacy policy? \textit{[exp1]}
    \begin{itemize}
        \item Almost daily / About once a week / About once a month / Less frequently / Never
    \end{itemize}
    \item How confident are you, if at all, that companies will do the following things?\\
    \indent \textit{The order of the following questions are randomized}
    \begin{enumerate}
        \item Follow what their privacy policies say they will do with your personal information \textit{[percep1]}
            \begin{itemize}
                \item Very confident / Somewhat confident / Not too confident / Not confident at all
            \end{itemize}
        \item Promptly notify you if your personal data has been misused or compromised \textit{[percep2]}
            \begin{itemize}
                \item Very confident / Somewhat confident / Not too confident / Not confident at all
            \end{itemize}
        \item Publicly admit mistakes and take responsibility when they misuse or compromise their
users’ personal data \textit{[percep3]}
            \begin{itemize}
                \item Very confident / Somewhat confident / Not too confident / Not confident at all
            \end{itemize}
        \item Use your personal information in ways you will feel comfortable with \textit{[percep4]}
            \begin{itemize}
                \item Very confident / Somewhat confident / Not too confident / Not confident at all
            \end{itemize}
        \item Be held accountable by the government if they misuse or compromise your data \textit{[percep5]}
            \begin{itemize}
                \item Very confident / Somewhat confident / Not too confident / Not confident at all
            \end{itemize}
    \end{enumerate}
    \item How comfortable are you, if at all, with companies using your personal data in the following ways?\\
    \indent \textit{The order of the first, second, and last questions are randomized}
    \begin{enumerate}
        \item To help improve their fraud prevention systems \textit{[percep6]}
            \begin{itemize}
                \item Very comfortable / Somewhat comfortable / Not too comfortable / Not comfortable at all
            \end{itemize}
        \item Sharing it with outside groups doing research that might help improve society \textit{[percep7]}
            \begin{itemize}
                \item Very comfortable / Somewhat comfortable / Not too comfortable / Not comfortable at all
            \end{itemize}
        \item This question is not part of the survey and just helps us to detect bots and automated scripts. To confirm that you are a human, please choose `Strongly agree' here.
            \begin{itemize}
                \item Strongly disagree / Disagree / Somewhat disagree / Neither agree nor disagree / Somewhat agree / Agree / Strongly Agree
            \end{itemize}
        \pagebreak
        \item To help them develop new products \textit{[percep8]}
            \begin{itemize}
                \item Very comfortable / Somewhat comfortable / Not too comfortable / Not comfortable at all
            \end{itemize}
    \end{enumerate}
    \item Have you recently decided NOT to use a product or service because you were worried about how much personal information would be collected about you? \textit{[behav1]}
    \begin{itemize}
        \item Yes, have done this / No, have not done this
    \end{itemize}
    \item Here’s a different kind of question. (If you don’t know the answer, select ``Not sure.'') As far as you know…
    \indent \textit{The order of the following questions is randomized. For each question, the order of the first four options is randomized.}
    \begin{enumerate}
        \item If a website uses cookies, it means that the site… \textit{[know1]}\\

        \begin{itemize}
            \item Can see the content of all the files on the device you are using 
            \item Is not a risk to infect your device with a computer virus 
            \item Will automatically prompt you to update your web browser software if it is out of date
            \item Can track your visits and activity on the site \textit{[correct]} 
            \item Not sure
        \end{itemize}
        \item Which of the following is the largest source of revenue for most major social media platforms? \textit{[know2]}\\
        \begin{itemize}
            \item Exclusive licensing deals with internet service providers and cellphone manufacturers
            \item  Allowing companies to purchase advertisements on their platforms \textit{[correct]}
            \item  Hosting conferences for social media influencers
            \item Providing consulting services to corporate clients
            \item Not sure
        \end{itemize}
        \item When a website has a privacy policy, it means that the site… \textit{[know3]}\\
        \begin{itemize}
            \item Has created a contract between itself and its users about how it will use their data \textit{[correct]} 
            \item Will not share its users’ personal information with third parties
            \item Adheres to federal guidelines about deceptive advertising practices 
            \item Does not retain any personally identifying information about its users 
            \item Not sure
        \end{itemize}
        \item What does it mean when a website has “https://” at the beginning of its URL, as opposed to “http://”without the “s”? \textit{[know4]}\\
        \begin{itemize}
            \item Information entered into the site is encrypted \textit{[correct]} 
            \item The content on the site is safe for children 
            \item The site is only accessible to people in certain countries 
            \item The site has been verified as trustworthy 
            \item Not sure
        \end{itemize}
        \item Where might someone encounter a phishing scam? \textit{[know5]}\\
        \begin{itemize}
            \item In an email 
            \item On social media 
            \item In a text message 
            \item On a website 
            \item All of the above \textit{[correct]} 
            \item None of the above 
            \item Not sure
        \end{itemize}
        \item Which two companies listed below are both owned by Facebook? \textit{[know6]}\\
        \begin{itemize}
            \item Twitter and Instagram 
            \item Snapchat and WhatsApp 
            \item WhatsApp and Instagram \textit{[correct]} 
            \item Twitter and Snapchat 
            \item Not sure
        \end{itemize}
        \item The term “net neutrality” describes the principle that… \textit{[know7]}\\
        \begin{itemize}
            \item Internet service providers should treat all traffic on their networks equally \textit{[correct]} 
            \item Social media platforms must give equal visibility to conservative and liberal points of
view 
            \item Online advertisers cannot post ads for housing or jobs that are only visible to people of a certain race 
            \item The government cannot censor online speech 
            \item Not sure
        \end{itemize}
        \pagebreak
        \item Many web browsers offer a feature known as “private browsing” or “incognito mode.” If someone opens a webpage on their computer at work using incognito mode, which of the following groups will NOT be able to see their online activities? \textit{[know8]}\\
        \begin{itemize}
            \item The group that runs their company’s internal computer network 
            \item Their company’s internet service provider 
            \item A coworker who uses the same computer \textit{[correct]} 
            \item The websites they visit while in private browsing mode 
            \item Not sure
        \end{itemize}
    \end{enumerate}
    \item Do you think that ALL Americans should have the right to have the following information about themselves removed from public online search results?\\
    \indent \textit{The order of the following questions is randomized}
    \begin{enumerate}
        \item Data collected by law enforcement, such as criminal records or mugshots \textit{[belief1]}
        \begin{itemize}
            \item Yes, should be able to remove this from online searches / No, should not be able to remove this from online searches
        \end{itemize}
        \item Information about their employment history or work record \textit{[belief2]}
        \begin{itemize}
            \item Yes, should be able to remove this from online searches / No, should not be able to remove this from online searches
        \end{itemize}
        \item Negative media coverage \textit{[belief3]}
        \begin{itemize}
            \item Yes, should be able to remove this from online searches / No, should not be able to remove this from online searches
        \end{itemize}
        \item Potentially embarrassing photos or videos \textit{[belief4]}
        \begin{itemize}
            \item Yes, should be able to remove this from online searches / No, should not be able to remove this from online searches
        \end{itemize}
    \end{enumerate}
    \item Today it is possible to take personal data about people from many different sources – such as their purchasing and credit histories, their online browsing or search behaviors, or their public voting records – and combine them together to create detailed profiles of people’s potential interests and characteristics. Companies and other organizations use these profiles to offer targeted advertisements or special deals, or to assess how risky people might be as customers.
    Prior to today, how much had you heard or read about this concept? \textit{[exp5]}
    \begin{itemize}
        \item A lot / A little / Nothing at all
    \end{itemize}
    \item In the last 12 months, have you had someone…\\
    \indent \textit{The order of the following questions is randomized}
    \begin{enumerate}
        \item Put fraudulent charges on your debit or credit card \textit{[exp2]}
        \begin{itemize}
            \item Yes / No
        \end{itemize}
        \item Take over your social media or email account without your permission \textit{[exp3]}
        \begin{itemize}
            \item Yes / No
        \end{itemize}
        \item Attempt to open a line of credit or apply for a loan using your name \textit{[exp4]}
        \begin{itemize}
            \item Yes / No
        \end{itemize}
    \end{enumerate}
    \item What is your age?
    \begin{itemize}
        \item 18-29 
        \item 30-49 
        \item 50-64 
        \item 65+ 
    \end{itemize}
    \item What is your sex?
    \begin{itemize}
        \item Male 
        \item Female
    \end{itemize}
    \item Please indicate your highest level of education
    \begin{itemize}
        \item Less than high school
        \item High school graduate 
        \item Some college, no degree 
        \item Associate's degree 
        \item College graduate/some post grad 
        \item Postgraduate
    \end{itemize}
    \item Choose the race that you consider yourself to be\\
    \indent \textit{The first four options are presented in randomized order}
    \begin{itemize}
        \item White 
        \item Black or African American 
        \item Asian or Asian American 
        \item Mixed Race 
        \item Some other race
    \end{itemize}
\end{enumerate}

\section{Survey Response Summaries} \label{appendix:tables}

Table~\ref{tab:propTab1} and Table~\ref{tab:propTab2} summarize how our four datasets compare on each of the thirty individual questions. Responses are within $\pm 5\%$ Pew proportions are highlighted in green; responses are $\geq 10\%$ off from Pew proportions ar highlighted in orange.
%\vspace{12pt}

\clearpage
\begin{table}[t!]
\centering
\begin{tabular}{lllrrr}
  \hline
Q & Ans & Pew & Repr & Bal & MTurk \\ 
  \hline
behav1 & Yes & 51.6 & \colorbox{green!30}{51.7} & \colorbox{green!30}{51.2} & \colorbox{orange!30}{62.6} \\ 
behav2 & Yes & 71.9 & 77.4 & \colorbox{green!30}{72} & \colorbox{orange!30}{93.2} \\ 
behav3 & Yes & 38.0 & \colorbox{orange!30}{63.5} & \colorbox{orange!30}{76.5} & \colorbox{orange!30}{88.9} \\ 
behav4 & Yes & 22.6 & \colorbox{orange!30}{58.8} & \colorbox{orange!30}{63.6} & \colorbox{orange!30}{88.7} \\ 
behav5 & Yes & 29.0 & \colorbox{orange!30}{73.1} & \colorbox{orange!30}{82.9} & \colorbox{orange!30}{84.9} \\ 
  \hline
  exp1 & Daily & 25.2 & 31.9 & \colorbox{green!30}{28} & 33.8 \\ 
   & Weekly & 32.1 & \colorbox{green!30}{35.8} & \colorbox{green!30}{35.6} & \colorbox{green!30}{34.2} \\ 
   & Monthly & 24.3 & 19.2 & \colorbox{green!30}{22.5} & \colorbox{green!30}{23.1} \\ 
   & Less & 15.4 & \colorbox{green!30}{12.6} & \colorbox{green!30}{13.6} & 7.8 \\ 
   & Never & 3.0 & \colorbox{green!30}{0.5} & \colorbox{green!30}{0.2} & \colorbox{green!30}{1.1} \\ 
  exp2 & Yes & 21.4 & 16.2 & 14.9 & \colorbox{orange!30}{45.2} \\ 
  exp3 & Yes & 8.0 & \colorbox{green!30}{5.4} & \colorbox{green!30}{6.4} & \colorbox{orange!30}{47.8} \\ 
  exp4 & Yes & 6.1 & \colorbox{green!30}{3.4} & \colorbox{green!30}{2.8} & \colorbox{orange!30}{48.1} \\ 
  exp5 & A lot & 27.2 & \colorbox{green!30}{31.8} & \colorbox{green!30}{27.6} & \colorbox{orange!30}{40.6} \\ 
   & A little & 49.8 & 57 & \colorbox{orange!30}{60.5} & \colorbox{green!30}{53.9} \\ 
   & Nothing & 23.0 & \colorbox{orange!30}{11.2} & \colorbox{orange!30}{11.9} & \colorbox{orange!30}{5.5} \\ 
   \hline
  know1 & Correct & 62.6 & \colorbox{orange!30}{89.7} & \colorbox{orange!30}{86.6} & \colorbox{orange!30}{48.6} \\ 
   & Incorrect & 9.3 & 3.8 & 2.8 & \colorbox{orange!30}{38} \\ 
   & Not sure & 28.2 & \colorbox{orange!30}{6.5} & \colorbox{orange!30}{10.6} & \colorbox{orange!30}{13.4} \\ 
  know2 & Correct & 58.9 & \colorbox{orange!30}{80.5} & \colorbox{orange!30}{80} & 48.9 \\ 
   & Incorrect & 8.5 & \colorbox{green!30}{4.1} & 3.4 & \colorbox{orange!30}{36.1} \\ 
   & Not sure & 32.6 & \colorbox{orange!30}{15.4} & \colorbox{orange!30}{16.6} & \colorbox{orange!30}{15} \\ 
  know3 & Correct & 47.8 & \colorbox{orange!30}{68.8} & \colorbox{orange!30}{71} & \colorbox{green!30}{44.4} \\ 
   & Incorrect & 24.6 & 18.5 & 15.6 & \colorbox{orange!30}{40.9} \\ 
   & Not sure & 27.6 & \colorbox{orange!30}{12.8} & \colorbox{orange!30}{13.4} & \colorbox{orange!30}{14.8} \\ 
  know4 & Correct & 30.3 & \colorbox{orange!30}{56.2} & \colorbox{orange!30}{54.2} & 37.8 \\ 
   & Incorrect & 15.1 & \colorbox{green!30}{13} & 9.9 & \colorbox{orange!30}{41} \\ 
   & Not sure & 54.6 & \colorbox{orange!30}{30.8} & \colorbox{orange!30}{35.9} & \colorbox{orange!30}{21.2} \\ 
  know5 & Correct & 67.1 & 75.5 & 76.2 & \colorbox{orange!30}{31.6} \\ 
   & Incorrect & 17.6 & \colorbox{green!30}{20.9} & \colorbox{green!30}{18.9} & \colorbox{orange!30}{58} \\ 
   & Not sure & 15.3 & \colorbox{orange!30}{3.6} & \colorbox{orange!30}{4.9} & \colorbox{green!30}{10.4} \\ 
  know6 & Correct & 28.7 & \colorbox{orange!30}{67} & \colorbox{orange!30}{73.1} & \colorbox{orange!30}{64.7} \\ 
   & Incorrect & 21.9 & 12.4 & 12.6 & \colorbox{green!30}{26} \\ 
   & Not sure & 49.4 & \colorbox{orange!30}{20.6} & \colorbox{orange!30}{14.2} & \colorbox{orange!30}{9.2} \\ 
  know7 & Correct & 44.6 & \colorbox{orange!30}{59.9} & \colorbox{orange!30}{58.3} & \colorbox{green!30}{43.2} \\ 
   & Incorrect & 12.0 & \colorbox{green!30}{12.9} & \colorbox{green!30}{13.9} & \colorbox{orange!30}{38.4} \\ 
   & Not sure & 43.4 & \colorbox{orange!30}{27.2} & \colorbox{orange!30}{27.9} & \colorbox{orange!30}{18.4} \\ 
  know8 & Correct & 24.4 & \colorbox{orange!30}{37.1} & \colorbox{orange!30}{53.4} & \colorbox{green!30}{26.6} \\ 
   & Incorrect & 25.5 & \colorbox{orange!30}{38.5} & \colorbox{green!30}{27.5} & \colorbox{orange!30}{53} \\ 
   & Not sure & 50.1 & \colorbox{orange!30}{24.4} & \colorbox{orange!30}{19.1} & \colorbox{orange!30}{20.4} \\ 
   \hline
\end{tabular}
 \caption{Proportions of responses to each question for the full samples. \colorbox{green!30}{Green} responses are within $\pm 5\%$ Pew proportions,  \colorbox{orange!30}{orange} responses are $\geq 10\%$ of Pew proportions.}
      \label{tab:propTab1}
\end{table}
\begin{table}[t!]
\centering
\begin{tabular}{lllrrr}
  \hline
Q & Ans & Pew & Repr & Bal & MTurk \\ 
  \hline
percep1 & VC & 4.8 & \colorbox{green!30}{5.5} & \colorbox{green!30}{5.9} & \colorbox{orange!30}{26.9} \\ 
   & SC & 37.1 & \colorbox{green!30}{41.2} & 42.8 & 44.4 \\ 
   & NTC & 40.3 & \colorbox{green!30}{37.6} & \colorbox{green!30}{37.5} & \colorbox{orange!30}{21} \\ 
   & NCAA & 17.7 & \colorbox{green!30}{15.6} & \colorbox{green!30}{13.9} & 7.8 \\ 
  percep2 & VC & 5.1 & \colorbox{green!30}{4} & \colorbox{green!30}{3.5} & \colorbox{orange!30}{26} \\ 
   & SC & 29.6 & \colorbox{green!30}{34.5} & \colorbox{green!30}{32.6} & 38.4 \\ 
   & NTC & 40.6 & \colorbox{green!30}{40.6} & \colorbox{green!30}{40.5} & \colorbox{orange!30}{22.5} \\ 
   & NCAA & 24.8 & \colorbox{green!30}{20.9} & \colorbox{green!30}{23.4} & \colorbox{orange!30}{13.1} \\ 
percep3 & VC & 2.9 & \colorbox{green!30}{2.8} & \colorbox{green!30}{2} & \colorbox{orange!30}{22.8} \\ 
   & SC & 17.8 & \colorbox{green!30}{21} & \colorbox{green!30}{17} & \colorbox{orange!30}{39.1} \\ 
   & NTC & 46.4 & \colorbox{green!30}{47} & \colorbox{green!30}{49} & \colorbox{orange!30}{25.9} \\ 
   & NCAA & 32.9 & \colorbox{green!30}{29.2} & \colorbox{green!30}{32} & \colorbox{orange!30}{12.2} \\ 
percep4 & VC & 3.6 & \colorbox{green!30}{3.4} & \colorbox{green!30}{3.1} & \colorbox{orange!30}{26.8} \\ 
   & SC & 27.2 & \colorbox{green!30}{28} & \colorbox{green!30}{27.6} & \colorbox{orange!30}{41.6} \\ 
   & NTC & 47.1 & \colorbox{green!30}{46} & \colorbox{green!30}{48.5} & \colorbox{orange!30}{22.4} \\ 
   & NCAA & 22.1 & \colorbox{green!30}{22.6} & \colorbox{green!30}{20.8} & \colorbox{orange!30}{9.2} \\ 
percep5 & VC & 3.6 & \colorbox{green!30}{4.4} & \colorbox{green!30}{2.9} & \colorbox{orange!30}{22.8} \\ 
   & SC & 27.2 & 18.9 & 18.2 & \colorbox{orange!30}{40.4} \\ 
   & NTC & 47.1 & \colorbox{green!30}{42.2} & \colorbox{green!30}{42.4} & \colorbox{orange!30}{21.5} \\ 
   & NCAA & 22.1 & \colorbox{orange!30}{34.5} & \colorbox{orange!30}{36.5} & 15.4 \\ 
percep6 & VC & 10.4 & \colorbox{green!30}{9.9} & \colorbox{green!30}{8.4} & \colorbox{orange!30}{28.9} \\ 
   & SC & 47.0 & \colorbox{green!30}{49} & \colorbox{green!30}{49.8} & \colorbox{green!30}{44.5} \\ 
   & NTC & 28.5 & \colorbox{green!30}{29.5} & \colorbox{green!30}{31.5} & \colorbox{orange!30}{18.1} \\ 
   & NCAA & 14.1 & \colorbox{green!30}{11.6} & \colorbox{green!30}{10.4} & 8.5 \\ 
percep7 & VC & 5.7 & \colorbox{green!30}{6.2} & \colorbox{green!30}{4.5} & \colorbox{orange!30}{23.5} \\ 
   & SC & 30.2 & \colorbox{green!30}{31.4} & \colorbox{green!30}{34.1} & \colorbox{orange!30}{40.8} \\ 
   & NTC & 36.6 & \colorbox{green!30}{35.9} & \colorbox{green!30}{35.1} & \colorbox{orange!30}{23.6} \\ 
   & NCAA & 27.4 & \colorbox{green!30}{26.5} & \colorbox{green!30}{26.2} & \colorbox{orange!30}{12.1} \\ 
percep8 & VC & 8.1 & \colorbox{green!30}{6.6} & \colorbox{green!30}{5} & \colorbox{orange!30}{29.8} \\ 
   & SC & 42.1 & \colorbox{green!30}{38} & \colorbox{green!30}{39.5} & \colorbox{green!30}{42.8} \\ 
   & NTC & 31.3 & \colorbox{green!30}{33.2} & \colorbox{green!30}{32.8} & \colorbox{orange!30}{16.6} \\ 
   & NCAA & 18.5 & \colorbox{green!30}{22.1} & \colorbox{green!30}{22.8} & 10.9 \\ 
   \hline
  belief1 & Yes & 39.1 & \colorbox{green!30}{43.2} & \colorbox{green!30}{38.8} & \colorbox{orange!30}{71} \\ 
  belief2 & Yes & 67.3 & 76.9 & 74.4 & \colorbox{orange!30}{82.3} \\ 
  belief3 & Yes & 56.1 & \colorbox{orange!30}{43.5} & \colorbox{orange!30}{36.4} & \colorbox{orange!30}{67.2} \\ 
  belief4 & Yes & 84.9 & \colorbox{green!30}{85.1} & \colorbox{green!30}{82} & \colorbox{green!30}{83} \\
  \hline
\end{tabular}
 \caption{Proportions of responses to each question for the full samples. \colorbox{green!30}{Green} responses are within $\pm 5\%$ Pew proportions,  \colorbox{orange!30}{orange} responses are $\geq 10\%$ of Pew proportions.
 \vspace{54pt}}
      \label{tab:propTab2}
\end{table}
\clearpage

\end{document}